\documentclass[aps,prl,twocolumn, superscriptaddress]{revtex4-1}
\pdfoutput=1
\usepackage[latin9]{inputenc}
\usepackage{amstext}
\usepackage{graphicx}
\usepackage{amssymb}
\usepackage{esint}

\begin{document}

\title{Collective dynamics in optomechanical arrays}

\author{Georg Heinrich}

\affiliation{Institute for Theoretical Physics, Universit\"at Erlangen-N\"urnberg, Staudtstr.
7, 91058 Erlangen, Germany}

\author{Max Ludwig}

\affiliation{Institute for Theoretical Physics, Universit\"at Erlangen-N\"urnberg, Staudtstr.
7, 91058 Erlangen, Germany}

\author{Jiang Qian}

\affiliation{Department of Physics, Center for NanoScience, and Arnold Sommerfeld
Center for Theoretical Physics, Ludwig-Maximilians-Universit\"at M\"unchen,
Theresienstr. 37, D-80333 M\"unchen, Germany}

\author{Bj\"orn Kubala}

\affiliation{Institute for Theoretical Physics, Universit\"at Erlangen-N\"urnberg, Staudtstr.
7, 91058 Erlangen, Germany}

\author{Florian Marquardt}

\affiliation{Institute for Theoretical Physics, Universit\"at Erlangen-N\"urnberg, Staudtstr.
7, 91058 Erlangen, Germany}

\affiliation{Max Planck Institute for the Science of Light, G\"unter-Scharowsky-Straße
1/Bau 24, 91058 Erlangen, Germany}

\maketitle

\begin{figure}
\includegraphics[width=1\columnwidth]{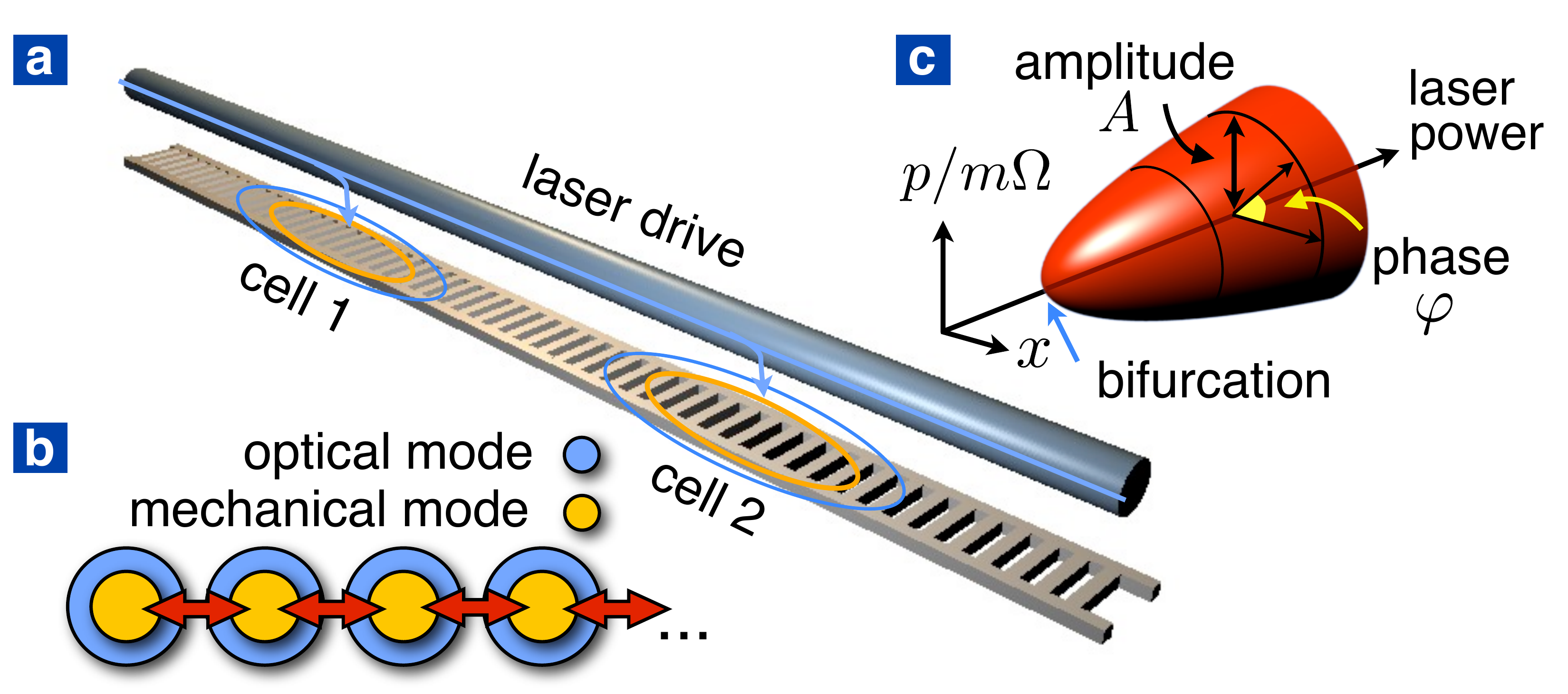}

\caption{\label{SetupFigure}Optomechanical crystals may be used to build arrays
with several localized optical and mechanical modes. (a) Potential
setup fabricated as a periodically patterned, free-standing dielectric
beam on a chip with laser drive via tapered fibre as in \cite{Eichenfield2009A-picogram--and,Eichenfield2009Optomechanical-}.
(b)\textbf{ }Schematic array of mechanically coupled optomechanical
cells. (c) For a single cell, at sufficient laser drive power, there
is a Hopf bifurcation towards self-induced oscillations with an undetermined
phase $\varphi$.}

\end{figure}

\textbf{The emerging field of optomechanics seeks to explore the interaction
between nanomechanics and light\textbf{ }(see \cite{Marquardt2009Optomechanics}
for a recent review). Rapid progress in laser cooling of nanomechanical
oscillators \cite{Schliesser2009Resolved-sideba,Groblacher2009Demonstration-o}
promises new fundamental tests of quantum mechanics \cite{Marshall2003Towards-Quantum},
while applications benefit from ultrasensitive detection of displacements,
masses and forces \cite{Anetsberger2009Near-field-cavi,Verlot2009Scheme-to-Probe,Teufel2009Nanomechanical-}.
Recently, the exciting concept of optomechanical crystals has been
introduced\textbf{ }\cite{Eichenfield2009A-picogram--and,Eichenfield2009Optomechanical-,Chang2010Slowing-and-sto},
where defects in photonic crystal structures are used to generate
both localized optical and mechanical modes that interact with each
other. For instance, this opens the prospect of integrated optomechanical
circuits combining several
functions on a single chip (see also \cite{Li2008Harnessing-opti,Li2009Tunable-bipolar}).
Here we start exploring the collective
dynamics of arrays consisting of many coupled optomechanical cells
(Fig.~\ref{SetupFigure}a,b). We show that such {}``optomechanical
arrays'' can display synchronization and that they can be described
by a modified Kuramoto model that allows to explain and predict most
of the features that will be observable in future experiments.}

The crucial ingredient of any optomechanical system is an optical
mode (OM) whose frequency shifts in response to a mechanical displacement:
$\delta\omega_{{\rm opt}}=-G x$. This coupling, vice versa,
gives rise to a radiation force, $F=\hbar G|\alpha|^{2}$, where $|\alpha|^{2}$
is the number of photons circulating inside the laser-driven OM. For
a laser red-detuned from the OM ($\Delta=\omega_{{\rm Laser}}-\omega_{{\rm opt}}<0)$,
dynamical back-action effects induced by the finite photon decay time
$\kappa^{-1}$ lead to cooling of the mechanical motion. In contrast,
for blue detuning ($\Delta>0$), anti-damping results. Once this overcomes
the internal mechanical friction, a Hopf bifurcation towards a regime
of self-induced mechanical oscillations takes place (Fig.~\ref{SetupFigure}c) 
\cite{Carmon2005Temporal-Behavi,Kippenberg2005Analysis-of-Rad,Marquardt2006Dynamical-Multi,Ludwig2008The-optomechani,Hohberger2004Self-oscillatio,Metzger2008Self-Induced-Os}.
While the mechanical amplitude $A$ is fixed, the oscillation phase
$\varphi$ is undetermined. Therefore, it is susceptible to external
perturbations. In particular, as we will see, it may lock to external
forces or to other optomechanical oscillators.

Synchronization has first been discovered by Huygens and
is now recognized as an important feature of collective nonequilibrium
behavior in fields ranging from physics over chemistry to biology
and neuroscience \cite{Pikovsky2001Synchronization},
with applications in signal processing and stabilization of oscillations.
A paradigmatic, widely studied model for synchronization was introduced
by Kuramoto \cite{Kuramoto1975Self-entrainmen}. For two oscillators,
his phase evolution equation reads $\dot{\varphi}_{1}=\Omega_{1}+K\sin(\varphi_{2}-\varphi_{1})$,
and likewise $\dot{\varphi}_{2}$. One finds synchronization ($\dot{\varphi}_{1}=\dot{\varphi}_{2}$)
if the coupling $K$ exceeds the threshold $K_{c}=|\Omega_{2}-\Omega_{1}|/2$, and the phase
lag $\delta\varphi=\varphi_{2}-\varphi_{1}$ vanishes for large $K$
according to $\sin(\delta\varphi)=(\Omega_{2}-\Omega_{1})/2K$. For
the globally coupled, mean-field type version of infinitely many oscillators,
there is a phase transition towards synchronization beyond some threshold
$K_{c}$ that depends on the frequency distribution \cite{Acebron2005The-Kuramoto-mo}.
In many examples the Kuramoto model is found as a generic, reduced
description of the phase dynamics. Nevertheless, for any specific system, it remains to be
seen whether this model (or possibly a structurally
similar variant thereof) applies at all, and how the coupling $K$
is connected to microscopic parameters \cite{Wiesenfeld1996Synchronization,Kozyreff2000Global-Coupling,Cross2004Synchronization}.
We now turn to this question in the case of optomechanical oscillators.

A single optomechanical cell consists of a mechanical mode (displacement
$x$) coupled to a laser-driven OM (light amplitude $\alpha$):
\begin{eqnarray}
m\ddot{x} & = & -m\Omega^{2}x-m\Gamma\dot{x}+\hbar G|\alpha|^{2}\label{eq:xevol}\\
\dot{\alpha} & = & \left[i(\Delta+Gx)-\frac{\kappa}{2}\right]\alpha+\frac{\kappa}{2}\alpha_{{\rm max}}\label{eq:alphaevol}\end{eqnarray}
Here $\Omega$ is the mechanical frequency, $\Gamma$ the intrinsic
damping, $G$ the optomechanical frequency pull per displacement,
and $\alpha_{{\rm max}}$ is the maximum light-field amplitude achieved
at resonance (set by the laser drive). 

Near the Hopf bifurcation (Fig.~\ref{SetupFigure}c), we can capture the essential dynamics by
eliminating the light field \cite{Marquardt2006Dynamical-Multi}\textbf{
}and switching to the phase- and amplitude-dynamics of the resulting Hopf oscillator:
\begin{eqnarray}
\dot{\varphi} & = & -\Omega+\frac{F(t)}{m\Omega A}\cos(\varphi)\label{eq:phaseevol}\\
\dot{A} & = & -\gamma(A-\bar{A})+\frac{F(t)}{m\Omega}\sin(\varphi).\label{eq:ampevol}\end{eqnarray}
Here $\bar{A}$ is the steady-state amplitude, and $\gamma$ is the
rate at which perturbations will relax back to $\bar{A}$. Both depend
on the microscopic optomechanical parameters, such as laser detuning and
laser drive power, and both vanish at the bifurcation threshold (see
methods section). Moreover, we have introduced an external force $F(t)$ (as added to Eq.~(\ref{eq:alphaevol}))

\begin{figure}
\includegraphics[width=1\columnwidth]{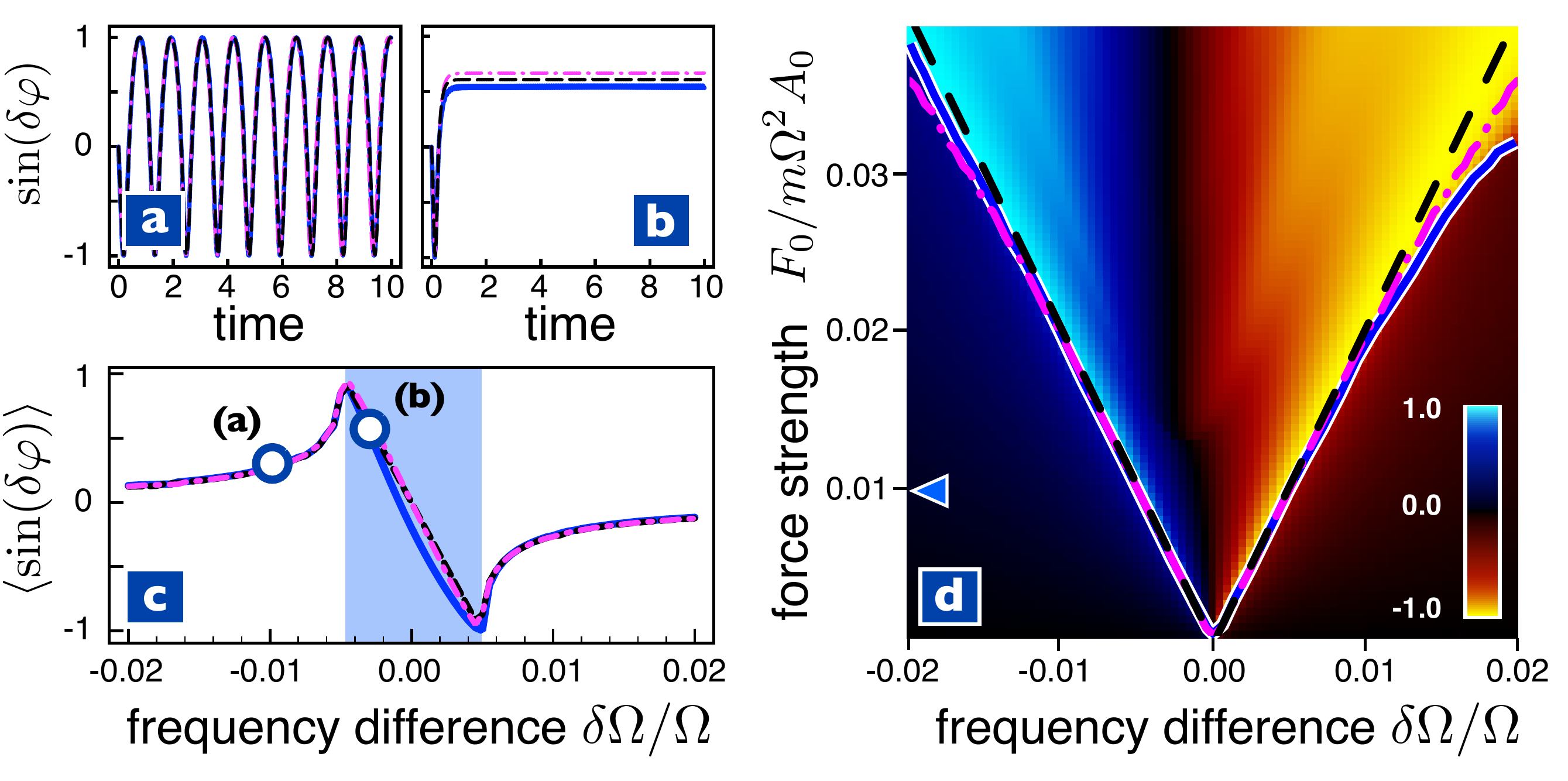}

\caption{\label{Fig2ExternalForce}Phase-locking of an optomechanical cell
to an external force. (a,b) Phase lag $\delta \varphi$ between oscillations and 
force, displayed via $\sin\delta\varphi$, outside
(a) and inside (b) the phase-locked regime (colored in (c)). Time
is plotted in units of $2\pi/\delta\Omega$. (c) Plot of $\left\langle \sin\delta\varphi\right\rangle $
as a function of frequency difference $\delta\Omega$, at the force strength
indicated by the blue triangle in (d), comparing optomechanics
(blue, solid) against the Hopf model (Eqs.~(\ref{eq:phaseevol},\ref{eq:ampevol});
magenta, dash-dot) and the Kuramoto model (Eq.~(\ref{eq:AdlerEq});
black, dash). (d) $\left\langle \sin\delta\varphi\right\rangle$
in the force vs. frequency difference plane, showing the Arnold tongue. Deviations between
the different models set in at high forces. (Lines indicate the transition
towards phase-locking, styles as in (c)).\textbf{ }(Microscopic cell
parameters: $\Delta/\Omega=1$, $\kappa/\Omega=1$, $\Gamma/\Omega=0.01$,
$\hbar G^{2}\alpha_{{\rm max}}^{2}/m\Omega^{3}=0.36$).}

\end{figure}

We start our discussion by considering phase-locking to an external force $F(t)=F_{0}\sin(\omega_F t)$. To this end we time-average Eq.~(\ref{eq:phaseevol}), keeping
only the slow dynamics, under the assumption $\omega_F\approx\Omega$.
This results in
\begin{equation}
\delta\dot{\varphi}=-\delta\Omega+K_{F}\sin(\delta\varphi)\,,\label{eq:AdlerEq}\end{equation}
where $\delta\varphi=\varphi(t)+\omega_F t$, $\delta\Omega=\Omega-\omega_F$,
and $K_{F}=F_{0}/2m\Omega\bar{A}$. Eq.~(\ref{eq:AdlerEq}) is a
special case of the Kuramoto equation. Direct numerical simulation
confirms the good agreement between the microscopic optomechanical
dynamics and the simplified descriptions via Eqs.~(\ref{eq:phaseevol},\ref{eq:ampevol})
and (\ref{eq:AdlerEq}). In Fig.~\ref{Fig2ExternalForce} we show $\sin\delta\varphi(t)$ and its time-average $\left\langle \sin\delta\varphi(t)\right\rangle $,
with the phase $\varphi$ being extracted from the complex amplitude
of motion, $\beta\equiv x+i\dot{x}/\Omega$. Phase-locking sets in
when $\delta\dot{\varphi}=0$ has a solution, i.e. for $|\delta\Omega|\leq K_{F}$,
resulting in an {}``Arnold tongue'' (see Fig.~\ref{Fig2ExternalForce}d).

We now turn to the dynamics of coupled cells, each of which is described
by Eqs.~(\ref{eq:xevol},~\ref{eq:alphaevol}). To these equations,
we add a mechanical coupling, set by a spring constant $k$: $m\ddot{x}_{1}=\ldots+k(x_{2}-x_{1})$.
In the Hopf model, this yields a force $F_{1}=kA_{2}\cos(\varphi_{2})$
on the first oscillator (and vice versa). The case of optical coupling
will be mentioned further below.

In order to arrive at the time-averaged dynamics for the phase difference,
$\delta\varphi=\varphi_{2}-\varphi_{1}$, it is necessary to go further
than before, keeping $A(t)=\bar{A}+\delta A(t)$ in the phase equation,
and eliminating the amplitude dynamics to lowest order (see methods for the derivation; and \cite{Aronson1990Amplitude-respo,Matthews1991Dynamics-of-a-l} for further examples where the amplitude dynamics is crucial).
Then, we arrive at an effective Kuramoto-type model for coupled optomechanical
Hopf oscillators:
\begin{equation}
\delta\dot{\varphi}=-\delta\Omega-C\cos(\delta\varphi)-K\sin(2\delta\varphi).\label{eq:Kuramoto}\end{equation}
In contrast to the standard Kuramoto model, $2\delta\varphi$ appears,
which will lead to both in-phase and anti-phase synchronization. This
corresponds to two distinct minima in the effective potential that
can be used to rewrite Eq.~(\ref{eq:Kuramoto}): $\delta\dot{\varphi}=-U'(\delta\varphi)$.
The coupling $K=k^{2}/2m^{2}\Omega^{2}\gamma$
diverges near the bifurcation, where $\gamma\rightarrow0$.
In the following we focus on the case of nearly
identical cells where the coupling $C$ can be neglected;
$C/\delta\Omega=k/2m\Omega^{2}\ll1$.

\begin{figure*}
\includegraphics[width=2\columnwidth]{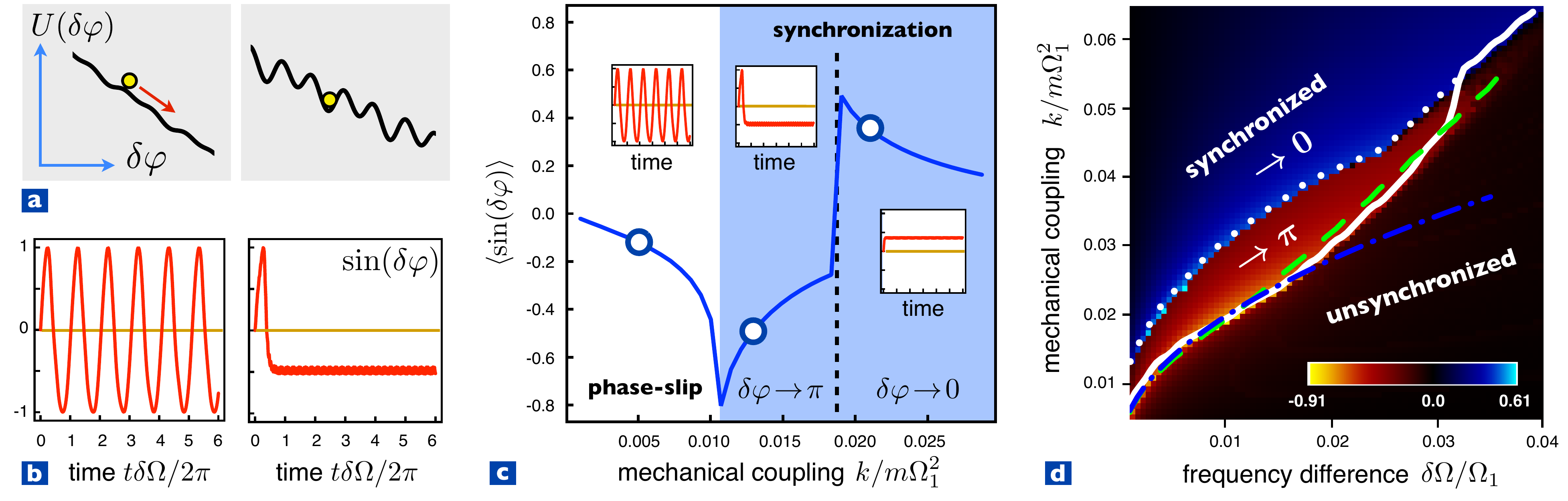}

\caption{\label{TwoCellsFig}Phase-locking for two coupled optomechanical cells. (a)
The phase particle in the effective Kuramoto potential $U(\delta\varphi)$,
in the de-synchronized and phase-locked regime
(left/right). (b) Time-evolution of $\sin(\delta\varphi(t))$. (c) When
the mechanical coupling $k$ exceeds a threshold, the phase difference
$\delta\varphi$ between the oscillations locks in spite of different
bare mechanical frequencies, here $\delta\Omega=0.003\,\Omega_{1}$. Both in-phase and
anti-phase regimes are observed.
(d) Phase lag, expressed via $\left\langle \sin(\delta\varphi)\right\rangle $,
in the plane coupling $k$ vs. frequency difference $\delta\Omega$,
including a comparison of the critical coupling $k_{c}$ (white, solid)
with the one from a Hopf model with one fit parameter (green,
dash) and the generalized Kuramoto model (blue, dash-dot). (Cell parameters
as in Fig.~\ref{Fig2ExternalForce}).}

\end{figure*}

To test whether these features are observed in the full optomechanical system,
we directly simulate the motion and increase the coupling $k$ for
a fixed frequency difference $\delta\Omega=\Omega_{2}-\Omega_{1}$.
The results are displayed in Fig.~\ref{TwoCellsFig}(a-c). Beyond
a threshold $k_{c}$, the frequencies and the phases lock, indicated
by a kink in $\left\langle \sin\delta\varphi\right\rangle $. As the
coupling increases further, the phases are pulled towards each other
even more, so $\left|\delta\varphi\right|$ decreases. Thus, coupled
optomechanical systems do indeed exhibit synchronization. As predicted,
there is both synchronization towards $\delta\varphi\rightarrow0$
and $\delta\varphi\rightarrow\pi$.

The dependence of the threshold $k_{c}$ on the frequency difference
$\delta\Omega$ is shown in Fig.~\ref{TwoCellsFig}d.
The observed behavior $k_{c} \sim \sqrt{\delta\Omega}$
at small $\delta\Omega$ is correctly reproduced by the generalized Kuramoto
model, Eq.~(\ref{eq:Kuramoto}).
For $\delta\Omega>\gamma$ deviations occur
via terms of higher order in $\delta\Omega/\gamma$, starting with
$ $$-(\delta\Omega/\gamma)K\cos(2\delta\varphi)$ in Eq.~(\ref{eq:Kuramoto}).
These produce a linear slope $k_{c}\propto\delta\Omega$, see Fig.~\ref{TwoCellsFig}d.

In terms of experimental realization, optomechanical crystals \cite{Eichenfield2009A-picogram--and,Eichenfield2009Optomechanical-}
offer a novel promising way to build arrays of optomechanical cells.
They are fabricated as free-standing photonic crystal beams (Fig.~\ref{SetupFigure}a).
Variations of the $\mu m$-scale lattice spacing produce both localized
optical and vibrational modes. The strong confinement leads to extremely
large optomechanical couplings, on the order of $G\sim100\,{\rm GHz/nm}$.
Typical parameters, that we use in our simulations for experimentally
realistic results, are $G=\text{100 GHz/nm}$, mechanical frequency
$\Omega=\text{1 GHz}$, mass $m=\text{100 fg}$, mechanical quality
factor $Q_{M}=\Omega/\Gamma=100$, cavity decay rate $\kappa=\text{1 GHz}$
and laser input powers such that the circulating photon number $|\alpha_{\text{max}}|^{2}\gtrsim100$. 

\begin{figure}
\includegraphics[width=1\columnwidth]{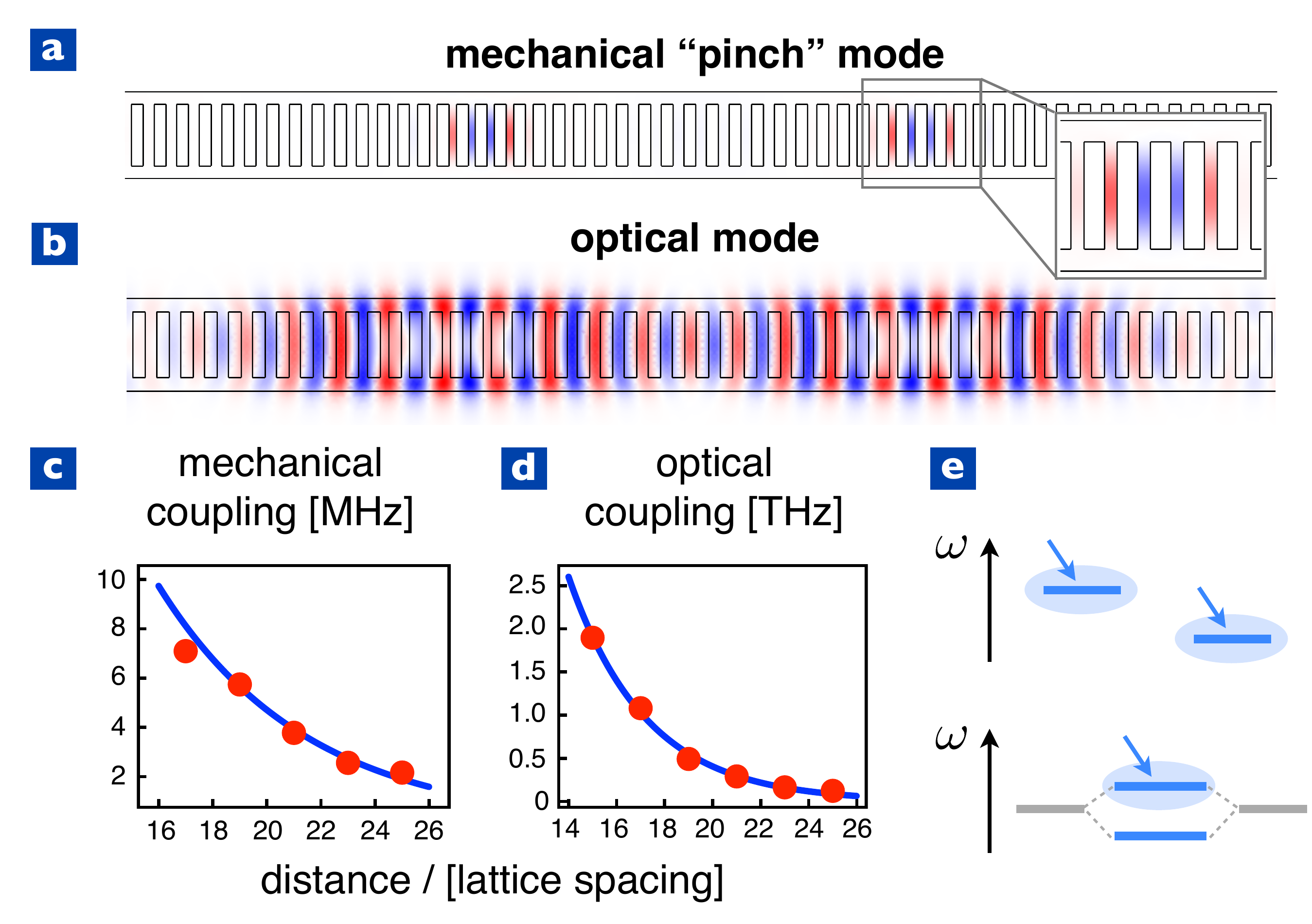}

\caption{\label{fig:FEMresults}FEM simulation for two coupled cells localized
on the same beam in an optomechanical crystal setup (cf. Fig.~\ref{SetupFigure}a).
(a) Horizontal displacement of vibrational {}``pinch'' modes. (b)
Transverse electric fields of optical modes. (c,d) The inter-cell
couplings decay exponentially as a function of distance. (e) Blue-detuned
lasers could illuminate two distinct optical modes (top) or a single
hybridized {}``molecular'' mode. (Geometrical parameters see methods
section).}

\end{figure}

To consider optomechanical arrays like in Fig.~\ref{SetupFigure}a,
we use finite element methods (FEM) to simulate two identical cells
arranged on the same beam (Fig.~\ref{fig:FEMresults}a,b). The optical
and vibrational couplings mediated by the beam can be deduced from
the splitting between the resulting symmetric/antisymmetric modes.
The results shown in Fig.~\ref{fig:FEMresults}c,d validate
the parameters considered above and indicate mechanical couplings $k/m\Omega^2 \lesssim 0.01$.
Due to the relatively strong optical coupling
($\sim\text{THz}$), distinct OMs in the individual cells
can only be achieved by patterning them to have frequencies sufficiently different
to prevent hybridization (Fig,~\ref{fig:FEMresults}e). This requires different
laser colors to address each cell. 

\begin{figure*}
\includegraphics[width=2\columnwidth]{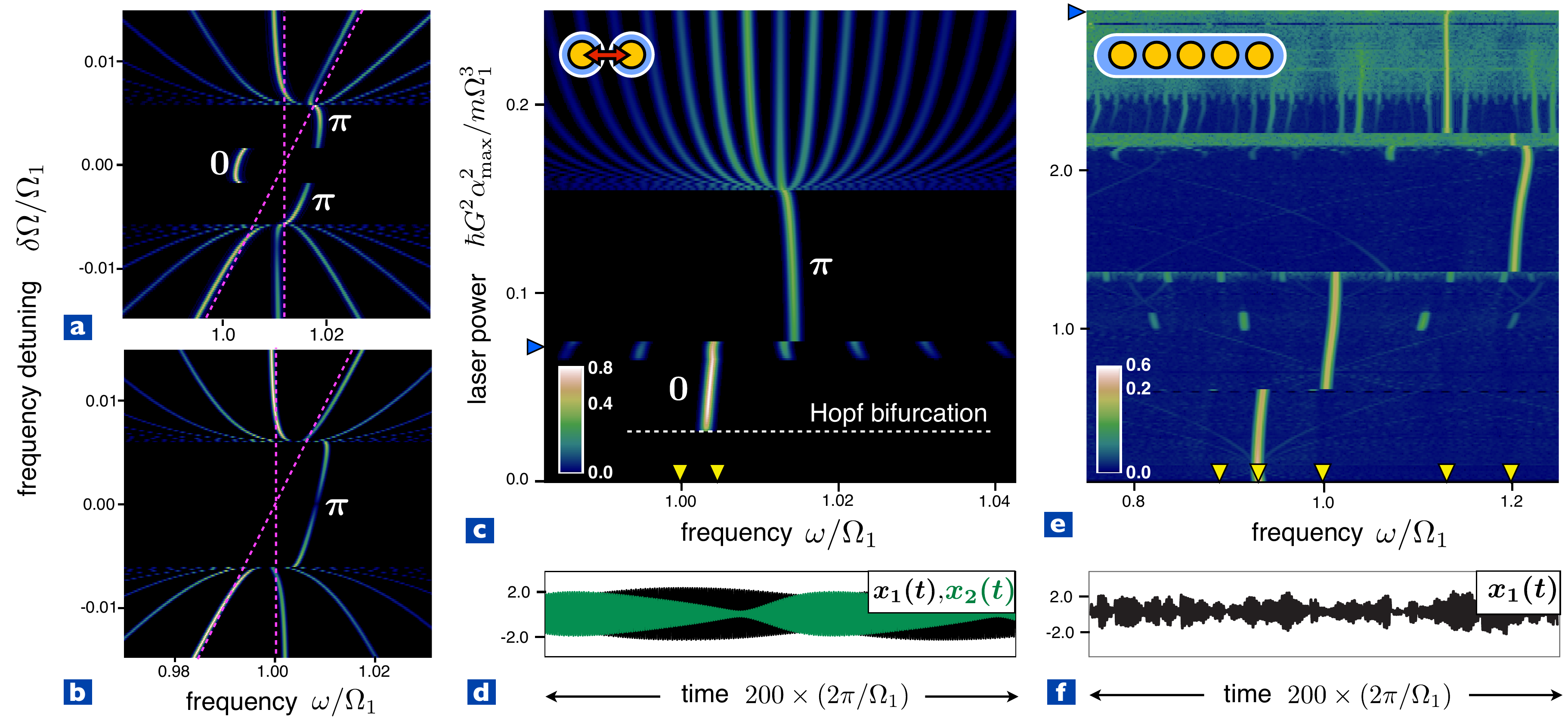}

\caption{\label{fig:ArraySynchro}Mechanical frequency spectra of optomechanical
arrays. (a,b) Frequency locking upon changing the detuning $\delta\Omega=\Omega_{2}-\Omega_{1}$
between the mechanical frequencies (magenta, dashed) of two independently
driven, mechanically coupled optomechanical cells ($k/m\Omega_{1}^{2}=0.015$).
(a) Spectrum $I(\omega)$ of intensity fluctuations, $I(t)=|\alpha_{1}(t)|^{2}+|\alpha_{2}(t)|^{2}$,
from an optomechanical model. (b) Spectrum of $x_{1}+x_{2}$ in the
Hopf model. (c) Spectrum vs. laser input power for two independently
driven, mechanically coupled cells ($k/m\Omega_{1}^{2}=0.01$, $\delta\Omega=0.005\Omega_{1}$).
(d) Example of trajectories $x_{i}(t)$, in units of $\kappa/G$, displaying
strong amplitude modulation, not described by the Kuramoto model (at
a power indicated by the blue triangle in (c)). (e) Spectrum $I(\omega)$
vs. laser input power for an array of five mechanical modes ($k=0$),
coupled to a common optical mode. At large drive, a regime of chaotic
motion is entered. Due to the presence of multiple attractors, the
regimes observed in such a diagram may depend on how the parameters
are swept. (f) Example of a trajectory in the chaotic regime (blue
triangle in (e)). (color scale in all plots indicates $|I(\omega)|$
in units of the peak height at $\omega=0$ for a system at rest; $\delta$
peaks are broadened for clarity; cell parameters are $\Delta_{i}=\kappa_{i}=100\Gamma_{i}=\Omega_{1}$,
as in Figs.~\ref{Fig2ExternalForce},\ref{TwoCellsFig}; yellow triangles
indicate the bare mechanical frequencies $\Omega_{i}$)}

\end{figure*}

In experiments, a convenient observable would be the RF frequency
spectrum of the light intensity emanating from the cells, $\left|\alpha\right|^{2}(\omega)$.
We first show the spectrum as a function of frequency difference $\delta\Omega$
for two mechanically coupled cells, driven independently (Fig.~\ref{fig:ArraySynchro}a).
Frequency locking is observed in an interval around $\delta\Omega=0$.
Experimentally, the mechanical frequencies can be tuned
via the {}``optical spring effect'' \cite{Eichenfield2009A-picogram--and}. 

The most easily tunable parameter is the laser drive power ($\propto\alpha_{{\rm max}}^{2}$).
Synchronization sets in right at the Hopf bifurcation.
For two cells (Fig.~\ref{fig:ArraySynchro}c), we recover the regimes of in-phase
and anti-phase synchronization. They differ in the synchronization frequency, 
$\bar{\Omega}(\pi)-\bar{\Omega}(0)=k/m\Omega$.
At higher drive, we find a transition towards de-synchronization.
This remarkable behavior can be explained from our analytical
results. We know that $\gamma$ increases away from the Hopf bifurcation
(i.e. for higher drive), leading to a concomitant decrease in the
effective Kuramoto coupling $K\propto1/\gamma$, and finally a loss
of synchronization. Near the transition, the frequencies fan out (as
$\delta\Omega_{{\rm eff}}\propto\sqrt{\alpha_{{\rm max}}-\alpha_{c}}$).
The multitude of peaks is produced due to nonlinear mixing. In some regimes, the
Kuramoto model fails (Fig.~\ref{fig:ArraySynchro}d).

For large arrays, it will be most practical to have identical OMs that combine
into extended 'molecular' modes, one of which is then driven by a
single laser via an evanescently coupled tapered fibre (Fig.~\ref{SetupFigure}a,
Fig.~\ref{fig:FEMresults}e). For efficient excitation of self-induced
oscillations, one has to ensure that the detuning $\Delta$ is equal
to all the mechanical frequencies $\Omega_{j}$ in the array, to within
$\left|\Delta-\Omega_{j}\right|<\kappa$. For arrays of reasonable
size, the splittings between adjacent optical molecular modes will
be more than $10^{2}$ times larger than $\kappa$, such that we can
ignore all but one OM. This setup then leads to a global coupling
of many nanoresonators to a single OM, such that

\begin{equation}
\dot{\alpha}=\left[i(\Delta+\sum_{j}G_{j}x_{j})-\frac{\kappa}{2}\right]\alpha+\frac{\kappa}{2}\alpha_{{\rm max}}\,,\label{eq:SingleOM}\end{equation}
and the force on each resonator is given by $-\hbar G_{j}|\alpha|^{2}$.
This setup comes close to realizing the all-to-all coupling most often
investigated in the literature on the Kuramoto model.

For illustration, we chose $N=5$ cells (Fig.~\ref{fig:ArraySynchro}e).
As before, we find synchronization regimes. In addition, at higher
drive, a transition into chaos takes place. Analyzing the time-evolution
in more detail, we observe transient fluctuations in amplitude and
phase, with a strong sensitivity on changes in initial conditions (Fig.~\ref{fig:ArraySynchro}f).
Note that in a single optomechanical cell one may also find chaotic
behavior \cite{Carmon2005Temporal-Behavi}, but for far larger driving
strengths.

Optomechanical arrays open up a new domain to study collective oscillator
dynamics, with room-temperature operation in integrated nanofabricated
circuits and with novel possibilities for readout and control, complementing
existing research on Josephson arrays \cite{Wiesenfeld1996Synchronization},
laser arrays \cite{Kozyreff2000Global-Coupling} and other nanomechanical
structures \cite{Cross2004Synchronization,Zalalutdinov2003Frequency-entra}.
Recent experiments on 2D optomechanical crystals \cite{Safavi-Naeini2010Optomechanics-i}
could form the basis for investigating collective dynamics in 2D settings
with various coupling schemes. Applications in signal processing may
benefit from phase noise suppression via synchronization \cite{Tallur2010Phase-Noise-Mod}.
Variations of the optomechanical arrays investigated here may also
be realized in other designs based on existing setups, like multiple
membranes in an optical cavity \cite{ThompsonStrong-dispersi}\textbf{
}or arrays of toroidal microcavities \cite{Schliesser2009Resolved-sideba,Anetsberger2009Near-field-cavi}.

\subsection*{Acknowledgements}
We thank O. Painter, H. Tang and J. Parpia for fruitful discussions, and GIF and DFG (Emmy-Noether program, NIM) for funding.

\subsection*{Methods}

The dynamics of a single optomechanical cell, Eqs.~(\ref{eq:xevol},~\ref{eq:alphaevol}),
in the self-induced oscillation regime can be mapped to a Hopf model
close to the Hopf bifurcation. This model is described by the steady
state amplitude $\bar{A}$ and amplitude decay rate $\gamma$ (see
Eqs.~(\ref{eq:phaseevol},~\ref{eq:ampevol})). The dependence of
$\gamma,\,\bar{A}$ on the microscopic parameters can be deduced by
expanding the average mechanical power input provided by the radiation
pressure force, $\hbar G\langle\left|\alpha\right|^{2}\dot{x}\rangle$
(see Eq.~(\ref{eq:xevol})), in terms of $a=G\bar{A}$; $\hbar G\langle\left|\alpha\right|^{2}\dot{x}\rangle=(\hbar \alpha_{{\rm max}}^2\pi_2)a^2+(\hbar \alpha_{{\rm max}}^2 \pi_4/\Omega^2)a^4+O(a^6)$. This yields
\[
\gamma=2\mathcal{P}\Omega\pi_{2}-\Gamma\]
 and \[
\left(G\bar{A}/\kappa\right)^{2}=\gamma\Omega/(-2\pi_{4}\mathcal{P}\kappa{}^{2}),\]
where the dimensionless coefficients $\pi_{2}(\Delta/\Omega,\kappa/\Omega)$
and $\pi_{4}(\Delta/\Omega,\kappa/\Omega)$ only depend on the rescaled
detuning and cavity decay rate. $\mathcal{P}=\hbar G^{2}\alpha_{{\rm max}}^{2}/m\Omega^{3}$
is the rescaled laser input power. Numerically, the Hopf parameters
may be found (even away from threshold) by simulating the exponential
relaxation of the cell's oscillation amplitude towards $\bar{A}$,
after a small instantaneous perturbation of the steady-state dynamics.
In general, to compare optomechanics to results from Hopf (e.g. Fig.~\ref{Fig2ExternalForce}),
the optical spring effect also needs to be considered.

Two mechanically coupled optomechanical cells are modeled as distinct
Hopf oscillators with phase dynamics $\varphi_{1}(t)$, $\varphi_{2}(t)$
and amplitude dynamics $A_{1}(t)$, $A_{2}(t)$ according to Eqs.~(\ref{eq:phaseevol})
and (\ref{eq:ampevol}). The coupling forces read $F_{1}=kA_{2}\cos(\varphi_{2})$,
$F_{2}=kA_{1}\cos(\varphi_{1})$. With $A_{i}(t)=\bar{A}_{i}+\delta A_{i}(t)$,
the solution for the amplitude dynamics is $\delta A_{i}(t)=\int_{-\infty}^{t}e^{-\gamma(t-t')}\tilde{f}_{i}(t')\, dt'$
where $\tilde{f_{1}}(t)=\frac{k(\bar{A}_{2}+\delta A_{2}(t))}{m_{1}\Omega_{1}}\cos\varphi_{2}(t)\sin\varphi_{1}(t)$
and likewise for $\tilde{f_{2}}(t)$. In the following, we consider
small couplings, where $\delta A_{i}\ll\bar{A}_{i}$. Then $\delta A_{1}$
(and likewise $\delta A_{2}$) is found to be \begin{eqnarray*}
\frac{\delta A_{1}(t)}{\bar{A}_{1}} & = & \frac{k}{2m_{1}\Omega_{1}}\times\\
 &  & \text{Im}\left(\frac{e^{i(\varphi_{2}(t)+\varphi_{1}(t))}}{\gamma-i(\Omega_{2}+\Omega_{1})}-\frac{e^{i(\varphi_{2}(t)-\varphi_{1}(t))}}{\gamma-i(\Omega_{2}-\Omega_{1})}\right).\end{eqnarray*}
Thus we can eliminate the amplitude dynamics to lowest order from
the phase equations, by expanding $A_{2}(t)/A_{1}(t)\approx\bar{A}_{2}/\bar{A}_{1}+\delta A_{2}/\bar{A}_{1}-\bar{A}_{2}\delta A_{1}/\bar{A}_{1}$
in the following equation (likewise for $\varphi_{2}$): \[
\dot{\varphi}_{1}=-\Omega_{1}+\frac{k}{m_{1}\Omega_{1}}\left(\frac{A_{2}}{A_{1}}\right)\cos(\varphi_{2})\cos(\varphi_{1}).\]
We now perform a time average, keeping only the slow dynamics near
frequencies $0$ and $\pm|\Omega_{2}-\Omega_{1}|$. This leads to
the stated result for the effective Kuramoto model, Eq.~(\ref{eq:Kuramoto}),
after setting $\delta\varphi=\varphi_{2}-\varphi_{1}$. The coupling
constants are given by $C=(k/2)\left(\bar{A}_{2}/m_{1}\Omega_{1}\bar{A}_{1}-\bar{A}_{1}/m_{2}\Omega_{2}\bar{A}_{2}\right)$
and $K=(1+(\xi_{1}/\xi_{2}+\xi_{2}/\xi_{1})/2)k^{2}/4m_{1}\Omega_{1}m_{2}\Omega_{2}$,
where $\xi_{j}=m_{j}\Omega_{j}\bar{A}_{j}^{2}$. 

The optomechanical simulation in Fig.~\ref{TwoCellsFig} shows results
for experimentally realistic microscopic parameters using an input
power well above the bifurcation threshold. This allows to observe
the essential features predicted from Hopf and the effective Kuramoto-type
model in an appropriate range of frequency detuning $\delta\Omega$.
However, to achieve quantitative agreement of the Hopf model in Fig.~\ref{TwoCellsFig},
its parameter $\gamma$ has to be treated as an adjustable parameter
(here $\gamma=0.02\,\Omega$). Each simulation initially starts with
a system at rest and considers an instantaneous switch-on of the laser
input power. Whether the system synchronizes towards $\delta\varphi\rightarrow0$
or $\delta\varphi\rightarrow\pi$ also depends on the initial conditions.

In principle, an amplitude dependence of the mechanical frequency,
$\Omega(A)\simeq\Omega(\bar{A})+(\partial\Omega/\partial A)\delta A$
can yield additional terms and an alternative synchronization mechanism
for two Hopf oscillators. However, numerical simulations verified
that for optomechanical cells this aspect can be neglected.

For the finite-element simulation in Fig.~\ref{fig:FEMresults} the
unit cell in the periodic part is a 1,396nm-wide, 362nm-long rectangle
with a co-centric rectangular hole of 992nm width and 190nm length.
The thickness of the beam is 220nm. The isotropic Young\textquoteright{}s
modulus of 168.5 GPa and the refractive index is 3.493. Each defect
is 15 units in length, symmetric across the 8th(central) cell. The
lattice constants vary linearly from 362nm at the edge to 307.7nm
at the center. The holes in the defects stay co-centric with the unit
cell and remain constant in size. For a single cell these parameters
have been reported in Ref.~\cite{Eichenfield2009Optomechanical-}.
\bibliographystyle{naturemag}
\bibliography{synchro}

\end{document}